\documentstyle[12pt, psfig]{article}
\def\reference{\parskip 0pt\par\noindent\hangindent 0.5 truecm}

\begin{document}
%
%
\title{Free--Free Absorption and the Unified Scheme}
%


\author{Seiji Kameno, $^{1}$ 
 Makoto Inoue, $^{1}$ 
 Kiyoaki Wajima, $^{1}$
 Satoko Sawada-Satoh $^{2}$, \and
 Zhi-Qiang Shen $^{2}$\thanks{Present address: Academia Sinica Institute of Astronomy and Astrophysics, Physics Departmen
t of National Taiwan University, No.1, Roosevelt Rd, Sec. 4, Taipei 106, Taiwan}
} 

\date{}
\maketitle

{\center
$^1$ National Astronomical Observatory of Japan, 2-21-1 Osawa, Mitaka Tokyo, Japan, 181-8588\\kameno@hotaka.mtk.nao.ac.jp\\inoue@nao.ac.jp\\kiyoaki@hotaka.mtk.nao.ac.jp\\[3mm]
$^2$ The Institute of Space and Astronautical Science, 3-1-1 Yoshinodai, Sagamihara Kanagawa, Japan, 229-8510\\satoko@vsop.isas.ac.jp\\zshen@asiaa.sinica.edu.tw\\
}

%
\begin{abstract}
We report Very-Long-Baseline Array (VLBA) observations at 2.3, 8.4, and 15.4 GHz towards nine GHz-Peaked Spectrum (GPS) sources.
One Seyfert 1 galaxy, one Seyfert 2 galaxy, three radio galaxies, and four quasars were included in our survey.
We obtained spatial distributions of the Free-Free Absorption (FFA) opacity with milliarcsec resolutions for all sources.
It is found that type-1 (Seyfert 1 and quasars) and type-2 (Seyfert 2 and radio galaxies) sources showed different distributions of the FFA opacities.
The type-1 sources tend to show more asymmetric opacity distributions towards a double lobe, while those of the type-2 sources are rather symmetric.
Our results imply that the different viewing angle of the jet causes the difference of FFA opacity along the external absorber.
This idea supports the unified scheme between quasars and radio galaxies, proposed by Barthel (1989).

\end{abstract}


\bigskip

%
%

\section{Introduction}
It is an important and controversial issue what causes the low-frequency cutoff in the radio spectrum of GPS sources; synchrotron self-absorption (SSA) or free--free absorption (FFA).
The discovery of FFA towards the GPS galaxy OQ 208 (Kameno et al. 2000)  propounded a question how general is FFA towards GPS sources.
The cold dense FFA plasma around the lobes of GPS sources could be a cocoon which smothers expansion of jets and lobes (Bicknell et al. 1997).

Such an external absorber can be a probe of the viewing angle of jets, so that a test for the unified scheme between quasars and radio galaxies (Barthel 1989), or the unified model between Seyfert 1 and 2 galaxies (Antonucci and Miller 1985) can be done.
Both models presume that these classes are intrinsically identical, and suggest that the apparent differences are due to the viewing angle.
With respect to Barthel's unified scheme, he showed that the projected distances of double lobes in quasars are significantly smaller than those of radio galaxies, and proposed that a smaller viewing angle causes higher luminosities and apparent presence of the broad line components.
On the Seyfert unification model, optical polarimetric observations for the Seyfert 2 galaxy NGC 1068 detected highly polarized continuum and broad Balmer lines (Miller and Antonucci 1983).
Antonucci and Miller (1985) suggested that hidden broad line components arose via scattering.
Presence of the broad line component, which is directly seen in Seyfert 1 but unseen in Seyfert 2 galaxies, was understood as intrinsic identity of the Seyfert 2 with Seyfert 1 galaxies.
Like the unification between quasars and radio galaxies, the viewing angle was thought to play a role in  different appearance.

When the jet axis is close to the line of sight, the path length through the external absorber will be longer towards the receding jet than towards the approaching jet.
Thus, the FFA opacity towards a double lobe is expected to be rather asymmetric; FFA opacity should be deeper towards the receding jet.
When the jet axis is nearly perpendicular to the line of sight, on the contrary, we expect rather symmetric FFA opacity towards a double lobe.

Based on this idea, we conducted a trichromatic GPS survey at 2.3, 8.4, and 15.4 GHz using the VLBA for 9 objects.

\section{The Sample}
We selected 9 sample objects, based on the GPS catalog by de Vries et al. (1997), under the criteria: 

\begin{enumerate}
\item The peak frequency $\nu_{\rm m}$ should stand within our observing range, i.e., $1.6$ GHz $< \nu_{\rm m} < 15$ GHz.
This condition is necessary to discriminate between SSA and FFA by spectral fitting, and then to obtain the distribution of opacities.

\item Since all sources must be bright enough to be detected with the VSOP and the VLBA, we put the criteria $S_{\rm 1.6 GHz} > 0.1$ Jy, $S_{\rm 5 GHz} > 0.5$ Jy, and $S_{\rm 15 GHz} > 0.2$ Jy.

\end{enumerate}

The selected GPS sources are listed in table \ref{tab:imageperform}.
The optical identifications are based on the NASA/IPAC Extragalactic Database (NED).
Hereafter, quasars and Seyfert 1 galaxies are categorized as type-1 sources, while radio galaxies and Seyfert 2 galaxies belong to type-2 sources. 

\begin{table}[t]
\caption{Trichromatic VLBA observations and image performance.}
\vspace{6pt}
{ \footnotesize
\begin{tabular*}{\textwidth}{@{\hspace{\tabcolsep} \extracolsep{\fill}}lp{6pc}crrrr} \hline \hline\\ [-6pt]
Object                                &
Frequency                             &
No. of                                & 
\multicolumn{3}{c}{Synthesized Beam}  & 
Image r. m. s.                        \\[4pt] \cline{4-6}
                                     &
 (GHz)                               &
Scans$^a$                            &
$\theta_{\rm maj}$ (mas)             &
$\theta_{\rm min}$ (mas)             &
p.a. ($^{\circ}$)                    &
(mJy/beam)                           \\[4pt] \hline
0108+388 & ~2.3~~\dotfill & 3 & $4.35$ & $2.35$ & $-12.1$ & $1.168$ \\
(RG)     & ~8.4~~\dotfill & 3 & $1.57$ & $0.77$ & $-10.1$ & $0.748$ \\
         & 15.4~~\dotfill & 7 & $0.81$ & $0.43$ & $-1.7$  & $0.482$ \\[6pt]
NGC 1052 & ~2.3~~\dotfill & 3 & $6.10$ & $2.52$ & $-4.4$  & $1.308$ \\
(Sy2)    & ~8.4~~\dotfill & 3 & $1.96$ & $0.83$ & $3.9$   & $1.118$ \\
         & 15.4~~\dotfill & 9 & $1.03$ & $0.40$ & $-2.9$  & $0.466$ \\[6pt]
0248+430 & ~2.3~~\dotfill & 3 & $5.32$ & $2.79$ & $-3.5$  & $0.772$ \\
(QSO)    & ~8.4~~\dotfill & 3 & $1.58$ & $0.85$ & $-7.7$  & $1.752$ \\
         & 15.4~~\dotfill & 9 & $0.66$ & $0.42$ & $-14.8$ & $0.478$ \\[6pt]
0646+600 & ~2.3~~\dotfill & 4 & $3.70$ & $1.95$ & $-20.8$ & $0.706$ \\
(QSO)    & ~8.4~~\dotfill & 4 & $1.35$ & $0.70$ & $-26.5$ & $0.604$ \\
         & 15.4~~\dotfill & 8 & $0.62$ & $0.40$ & $-42.7$ & $0.810$ \\[6pt]
0738+313 & ~2.3~~\dotfill & 3 & $5.55$ & $2.26$ & $-14.2$ & $3.741$ \\
(QSO)    & ~8.4~~\dotfill & 3 & $1.89$ & $0.76$ & $-13.6$ & $1.547$ \\
         & 15.4~~\dotfill & 9 & $0.81$ & $0.41$ & $-11.5$ & $2.696$ \\[6pt]
1333+459 & ~2.3~~\dotfill & 5 & $4.44$ & $2.02$ & $-1.7$  & $0.573$ \\
(QSO)    & ~8.4~~\dotfill & 5 & $1.55$ & $0.70$ & $-0.8$  & $0.553$ \\
         & 15.4~~\dotfill & 9 & $0.82$ & $0.40$ & $-12.8$ & $0.711$ \\[6pt]
1843+356 & ~2.3~~\dotfill & 3 & $5.02$ & $2.60$ & $6.7$   & $1.321$ \\
(RG)     & ~8.4~~\dotfill & 3 & $1.64$ & $0.84$ & $12.4$  & $1.049$ \\
         & 15.4~~\dotfill & 9 & $0.77$ & $0.42$ & $3.1$   & $0.528$ \\[6pt]
2050+364 & ~2.3~~\dotfill & 2 & $6.22$ & $2.82$ & $0.0$   & $2.293$ \\
(RG)     & ~8.4~~\dotfill & 2 & $1.82$ & $0.76$ & $-0.7$  & $3.545$ \\
         & 15.4~~\dotfill & 11 & $0.72$ & $0.44$ & $-5.5$  & $0.375$ \\[6pt]
2149+056 & ~2.3~~\dotfill & 2 & $6.80$ & $2.35$ & $-5.2$  & $0.572$ \\
(Sy1)    & ~8.4~~\dotfill & 2 & $1.89$ & $0.76$ & $-0.7$  & $0.493$ \\
         & 15.4~~\dotfill & 8 & $0.93$ & $0.44$ & $-1.6$  & $0.576$ \\[6pt]
\hline
\end{tabular*}
}
\\[4pt]
{\footnotesize $^a$One scan corresponds to integration of 11 minutes.}
\label{tab:imageperform}
\end{table}

\begin{table}[t]
\caption{Flux densities and FFA parameters of each component.}
\vspace{6pt}
{
\begin{tabular*}{\textwidth}{@{\hspace{\tabcolsep} \extracolsep{\fill}}p{6pc}cccc} \hline \hline\\ [-6pt]
Object \&       & Com- & \multicolumn{3}{c}{Flux density}  \\ \cline{3-5}
Optical ID$^*$    & ponent & 2.3 GHz     & 8.4 GHz      & 15.4 GHz     \\
                  &    & (mJy)       & (mJy)        & (mJy)        \\ \hline
0108+388 \dotfill & A  & ~591$\pm$~28 & ~574$\pm$~60 & ~262$\pm$~26  \\
(RG)              & B  & ~429$\pm$~21 & ~286$\pm$~25 & ~152$\pm$~4  \\
                  & C  & $\cdots$     & $\cdots$     & ~~16$\pm$~3  \\ [6pt]

NGC 1052 \dotfill & A  & ~~13$\pm$~~6 & ~247$\pm$~25 & ~189$\pm$18  \\
(Sy2)             & B  & ~969$\pm$~71 & 1796$\pm$184 & 1311$\pm$43  \\
                  & C  & ~~~~~$<$5.6 & ~~77$\pm$~51 & ~489$\pm$13  \\ [6pt]

0248+430 \dotfill & A  & 1051$\pm$~72 & ~884$\pm$~65 & ~668$\pm$14  \\ 
(QSO)             & B  & ~208$\pm$~~9 & ~~96$\pm$~~9 & ~~68$\pm$~2  \\
                  & C  & ~~88$\pm$~11 & ~~14$\pm$~~4 & ~~~5$\pm$~1  \\ [6pt]

0646+600 \dotfill & A  & ~360$\pm$~26 & ~882$\pm$~65 & ~866$\pm$16  \\
(QSO)             & B  & ~402$\pm$~25 & ~280$\pm$~24 & ~120$\pm$~8  \\
                  & C  & $\cdots$     & ~~12$\pm$~~2 & ~~11$\pm$~2  \\ [6pt]

0738+313 \dotfill & A  & ~187$\pm$~25 & ~981$\pm$133 & ~912$\pm$95  \\
(QSO)             & B  & 2635$\pm$131& 2627$\pm$228 & 2156$\pm$72  \\ [6pt]

1333+459 \dotfill & A  & ~168$\pm$~42 & ~319$\pm$~40 & ~252$\pm$~6  \\
(QSO)             & B  & ~372$\pm$~34 & ~204$\pm$~40 & ~~91$\pm$~4  \\ [6pt]

1843+356 \dotfill & A  & ~143$\pm$~~6 & ~221$\pm$~20 & ~119$\pm$~3  \\
(RG)              & B  & ~828$\pm$~55 & ~225$\pm$~16 & ~~98$\pm$~3  \\ [6pt]

2050+364 \dotfill & A  & 2458$\pm$~99 & 1114$\pm$~81 & ~559$\pm$11  \\
(RG)              & B  & 2996$\pm$121 & ~790$\pm$~59 & ~375$\pm$10  \\ [6pt]

2149+056 \dotfill & A  & ~771$\pm$~46 & ~572$\pm$~42 & ~393$\pm$~9  \\ 
(Sy1)             & B  & ~111$\pm$~92 & ~~47$\pm$~~7 & ~~28$\pm$~8  \\ [6pt] \hline
\end{tabular*}
}
\\[4pt]
{\footnotesize $^*$Optical identifications are based on the NASA/IPAC Extragalactic Database (NED). QSO, RG, and Sy stand for quasars, radio galaxies, and Seyfert galaxies, respectively.}
\label{tab:compflux}
\end{table}

\section{Observations and Results}
The VLBA observations have been carried out on December 15, 1998.
Table \ref{tab:imageperform} lists the sample objects of the VLBA observations.
Every object was observed at 3 frequencies with 2 to 11 scans, where each scan corresponds to integration of 11 minutes.
Observations at three frequencies have been carried out almost simultaneously.
The subreflector switched between the dual-frequency 2.3/8.4 GHz reflector system and the 15.4 GHz feed horn within a 1-minute gap between the cycles.
We used 4 channels of 8-MHz bandwidth at 15.4 GHz, and allocated 2 channels at both 2.3 and 8.4 GHz.
The correlation process was accomplished by the VLBA correlator.
We applied fringe-fitting, data flagging, and {\it a priori} amplitude calibration in the NRAO AIPS.
Imaging and self-calibration processes were carried out by Difmap.
Synthesized beam sizes and image qualities are listed in table \ref{tab:imageperform}.

While relative gain errors among the antennae are corrected though amplitude self-calibration processes, further flux calibration is necessary to obtain certain spectra across the observing frequencies. 
For the purpose of absolute flux calibration, we also imaged four calibrators; the BL Lacertae, DA 193, 3C 279, and OT 081.
These calibrators are so compact that the total CLEANed flux densities should be the same with the total flux densities measured by a reliable single-dish or a short-baseline-interferometer.
Basing on the comparison between the total flux measurements by the University of Michigan Radio Astronomy Observatory (UMRAO) and the NRAO Green Bank interferometer, and summation of CLEANed flux densities, we applied flux scaling for the final results.  
At 15.4 GHz, for instance, total CLEANed flux densities before absolute correction were 3.276, 5.012, 25.769, and 4.172 Jy for BL Lac, DA 193, 3C 279, and OT 081, respectively.
The UMRAO database provides the total flux densities at 14.5 GHz, $3.458 \pm 0.015$, $4.763 \pm 0.035$, $27.69 \pm 0.18$, and $4.380 \pm 0.030$ Jy for each, as averaged over 2 months centered at our observation.
Then we calculated a correction factor ($C = \frac{S_{\rm cc}}{S_{\rm U}}$), where $S_{\rm cc}$ is the total CLEANed flux density of the calibrator images, and $S_{\rm U}$ is the mean flux densities of calibrators measured by the UMRAO.
We have $C =  0.959 \pm 0.018$ at 15 GHz, so that we scaled our image by the factor and obtained the accuracy of the flux scale of 1.8 \%.
The accuracy was derived from a standard deviation of the correction factors by the four calibrators. 
In the same manner, we scaled images at 2.3 and 8.4 GHz by the correction factors of 0.84 and 0.79, respectively.
The estimated amplitude accuracies are 4.0, 7.2 and 1.8 \% at 2.3, 8.4, and 15.4 GHz, respectively.

\begin{figure}[htb]
\centerline{
\psfig{file=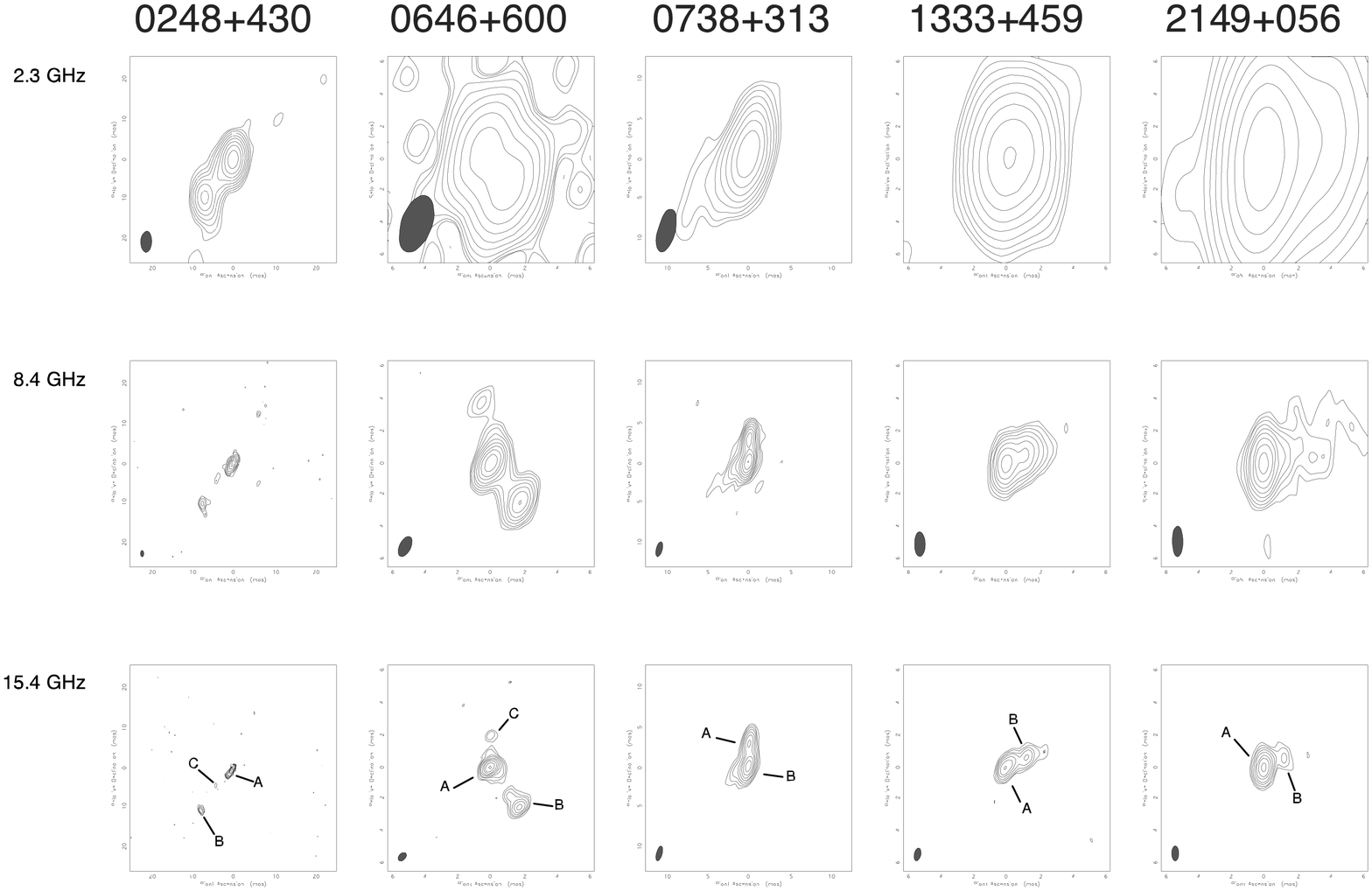,width=132mm}
}
\bigskip
\bigskip
\centerline{
\psfig{file=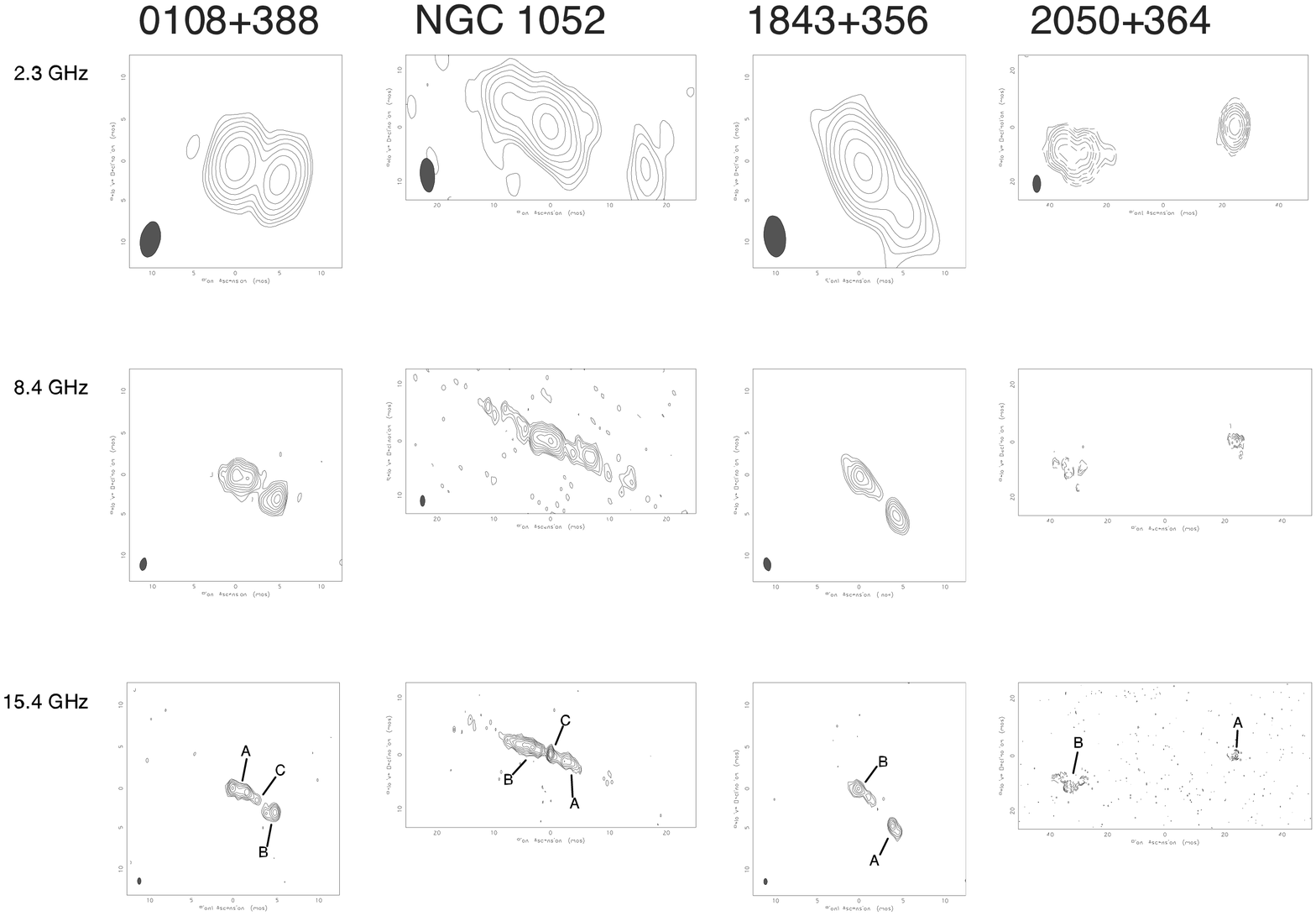,width=122.72mm}
}
\caption{Trichromatic images and distributions of FFA opacity of nine GPS sources. Five type-1 sources and four type-2 sources are shown in upper and lower panels, respectively. Contours start at $\pm 3 \sigma$ level, increasing by factors of 2. The FFA opacity $\tau_{\rm f}$ is derived by spectral fitting with equation \ref{eqn:ffaspectrum}.
}
\label{fig:images}            
\end{figure}

All images with uniform weighting are shown in figure \ref{fig:images}.
Based on the images, we estimated a flux density of each component using three different methods; tasks `IMFIT' and `JMFIT' to adopting elliptical Gaussian distribution of the brightness, and `TVSTAT' to integrate brightness within a specified area.
When the resolution is insufficient to isolate a double structure, in the cases of 2.3 GHz images of 0738+313, 1333+459, and 2149+056, we applied double elliptical Gaussian fits for `IMFIT' and 'JMFIT'.
In these cased, we also tried Gaussian model fits in visibilities using `modelfit' in Difmap instead of `TVSTAT', because these images are simple enough to be fitted in visibilities while it is difficult to set adequate integral area in these images for `TVSTAT'.
The flux densities of components are shown in table \ref{tab:compflux}.
Errors in the flux densities are estimated by the root sum squared of the amplitude calibration errors and the uncertainties in measurement of flux densities from the images.
The uncertainties in measurement are evaluated by the standard deviation between results of the three methods, i.e. `IMFIT', `JMFIT', and `TVSTAT' (or `modelfit').
Components which are considered to be lobes are labeled as A and B, in the sense that a component with a larger FFA opacity is labeled as A. 
The identification of lobes or a core is discussed in the next section.

Since we have observed at only three frequencies, it is impossible to discriminate between SSA and FFA by the spectral fit alone.
Instead, we started from the simple FFA model assuming that SSA is negligible, and verify the assumed condition later.
It should be noticed that a solid model fit to discriminate between SSA and FFA requires at least five frequencies.
We are conducting VSOP observations at 1.6 and 4.8 GHz for the supplement, which will be reported in continuation.
Nevertheless, global properties among type-1 and -2 GPS sources can be discussed as shown later.

Anyway, we attempted the model: an optically thin synchrotron emission from the lobe is absorbed by external FFA plasma,
\begin{eqnarray}
S_{\nu} = S_0 \nu^{\alpha_0} \exp(-\tau_{\rm f} \nu^{-2.1}). \label{eqn:ffaspectrum}
\end{eqnarray}
Here, $S_{\nu}$ is the observed flux density in Jy, $S_0$ is the intrinsic flux density in Jy at 1 GHz, $\nu$ is the frequency in GHz, $\alpha_0$ is the intrinsic spectral index of the synchrotron emission, and $\tau_{\rm f}$ is the FFA coefficient.
Free parameters are $S_0$ and $\tau_{\rm f}$ for each Gaussian component, and common $\alpha_0$ for all components.
Therefore, the number of free parameters $N_{\rm param} = 2n + 1$, where $n$ is the number of Gaussian components.
The number of data points is given by three frequencies times the number of Gaussian components, hence $3n$. Consequently, the degree of freedom will be $3n - N_{\rm param} = n-1$.
When we have two Gaussian components, the degree of freedom is 1.
Again, the number of frequencies is too few to verify a statistical confidence for the spectral fit.
Despite this condition, the flux densities, the opacity coefficients, and the residuals of spectral fits are listed in table \ref{tab:ffafitparam}.
Note that the FFA coefficient $\tau_{\rm f}$ directly corresponds to the spectral peak frequency $\nu_{\rm m}$ because the spectral peak appears at the frequency where the optical depth approximates to unity.
Thus, we have
\begin{eqnarray}
\nu_{\rm m} \simeq \tau_{\rm f}^{\frac{1}{2.1}}. \label{eqn:peakfreq}
\end{eqnarray}

\begin{figure}[htb]
\centerline{
\psfig{file=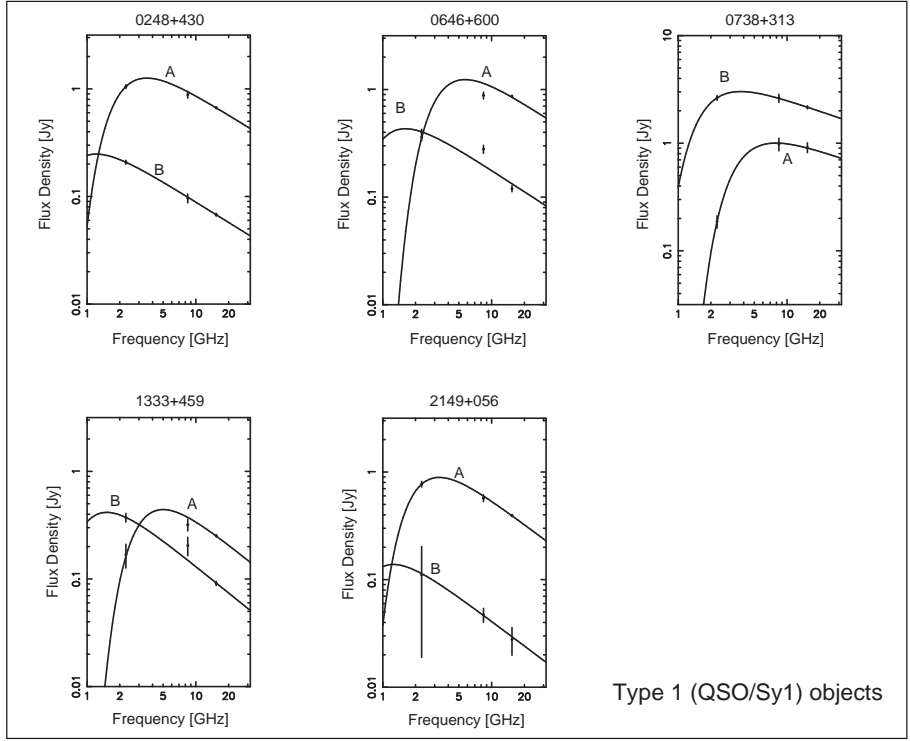,width=120mm}
}
\caption{Spectra of type-1 sources.
Flux densities at 2.3, 8.4, and 15.4 GHz are measured by Gaussian model fitting of CLEAN images using the task `IMFIT' in AIPS, as listed in table \ref{tab:compflux}.
Errors are RSS (root sum squares) of fitting errors shown in IMFIT and flux calibration errors.
The solid lines are results of the FFA spectral fitting with the model spectrum defined as equation \ref{eqn:ffaspectrum}.
Fitting parameters are listed in table \ref{tab:ffafitparam}.}
\label{fig:SPECtype1}            
\end{figure}

\begin{figure}[htb]
\begin{center}
\psfig{file=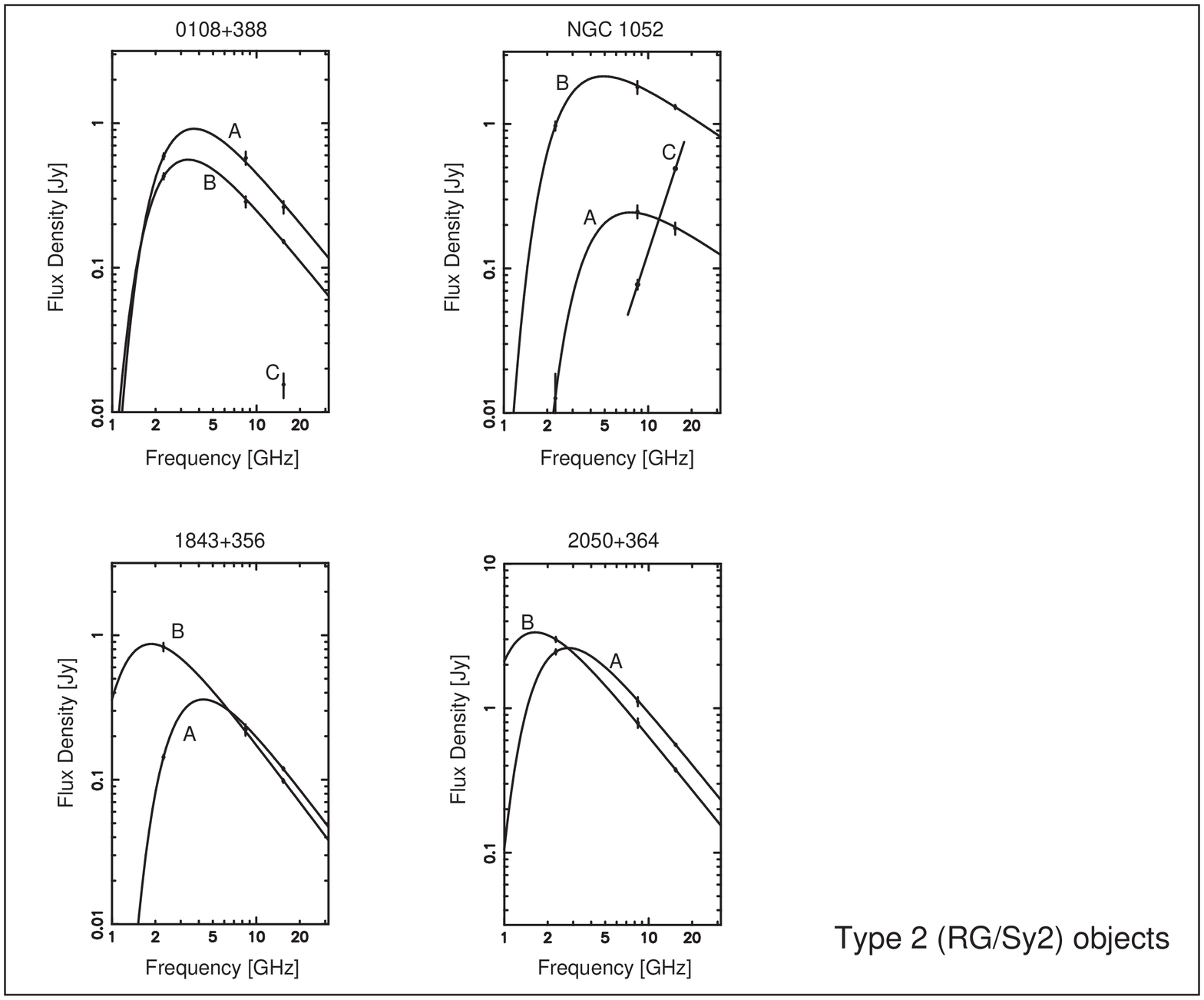,width=120mm}
\end{center}
\caption[Spectra of type-2 GPS sources.]{\footnotesize Spectra of type-2 sources.
Descriptions are the same with figure \ref{fig:SPECtype1}
}
\label{fig:SPECtype2}
\end{figure}

Derived best-fit spectra are shown in figures \ref{fig:SPECtype1} and \ref{fig:SPECtype2} for type-1 and -2 sources, respectively.
Figure \ref{fig:FFAhistogram} shows the histograms of FFA opacities towards each component.

\begin{table}[t]
\caption{Flux densities and FFA parameters of each component.}
\vspace{6pt}
{\small
\begin{tabular*}{\textwidth}{@{\hspace{\tabcolsep} \extracolsep{\fill}}p{8pc}ccccc} \hline \hline\\ [-6pt]
Object \&         & $\alpha_0$ & Com-   & $S_0$       & $\tau_{\rm f}$     & $\chi^2$  \\ 
Optical ID        &            & ponent & (Jy)        &              &     \\[4pt] \hline
0108+388 \dotfill & $-1.23$    & A      & ~8.18$\pm$0.61 & ~9.07$\pm$0.50 & 0.67 \\
(RG)              &            & B      & ~4.46$\pm$0.12 & ~7.46$\pm$0.32 & 0.22 \\[6pt]

NGC1052  \dotfill & $-0.69$    & A      & ~1.37$\pm$0.10 & 23.0$\pm$2.72  & 0.01 \\
(Sy2)             &            & B      & ~8.88$\pm$0.28 & ~9.26$\pm$0.46 & 0.05 \\[6pt]

0248+430 \dotfill & $-0.63$    & A      & ~3.77$\pm$0.08 & ~4.27$\pm$0.26 & 0.75 \\ 
(QSO)             &            & B      & ~0.38$\pm$0.01 & ~0.45$\pm$0.30 & 0.05 \\ [6pt]

0646+600 \dotfill & $-0.66$    & A      & ~5.34$\pm$0.10 & 12.2 $\pm$0.43 & 17.3 \\
(QSO)             &            & B      & ~0.81$\pm$0.04 & ~0.86$\pm$0.47 & 14.6 \\ [6pt]

0738+313 \dotfill & $-0.35$    & A      & ~2.44$\pm$0.21 & 12.8 $\pm$0.92 & 0.04 \\
(QSO)             &            & B      & ~5.66$\pm$0.18 & ~2.68$\pm$0.34 & 0.01 \\ [6pt]

1333+459 \dotfill & $-0.82$    & A      & ~2.44$\pm$0.06 & 11.5 $\pm$1.48 & 1.90 \\
(QSO)             &            & B      & ~0.86$\pm$0.04 & ~0.94$\pm$0.58 & 1.96 \\ [6pt]

1843+356 \dotfill & $-1.33$    & A      & ~4.68$\pm$0.12 & 13.5 $\pm$0.24 & 0.58 \\
(RG)              &            & B      & ~3.76$\pm$0.11 & ~2.35$\pm$0.41 & 0.45 \\ [6pt]

2050+364 \dotfill & $-1.24$    & A      & 16.81$\pm$0.33 & ~5.08$\pm$0.25 & 0.04 \\
(RG)              &            & B      & 11.16$\pm$0.29 & ~1.66$\pm$0.27 & 0.04 \\ [6pt]

2149+056 \dotfill & $-0.76$    & A      & ~3.17$\pm$0.06 & ~4.44$\pm$0.36 & 0.36 \\ 
(Sy1)             &            & B      & ~0.24$\pm$0.03 & ~0.58$\pm$4.82 & 0.06 \\ [6pt]
\hline
\end{tabular*}
}
\label{tab:ffafitparam}
\end{table}

\section{Discussion}

A significant difference between type-1 and -2 GPS sources can be seen in the FFA fitting.
Spectral peak frequencies of double components in type-1 sources tend to differ significantly, while those in type-2 sources are relatively similar.  
In other words, type-1 sources tend to show asymmetric FFA opacities towards double lobes, while type-2 sources have rather symmetric opacities, since the $\nu_{\rm m}$ and $\tau_{\rm f}$ are related by equation \ref{eqn:peakfreq}.
To evaluate the asymmetry in the FFA opacities, we define the FFA opacity ratio $R=\tau_{\rm fA} / \tau_{\rm fB}$ ($\tau_{\rm fA} > \tau_{\rm fB}$).
$R$ is an index of asymmetry in the peak frequencies $\nu_{\rm m}$, too, as
\begin{eqnarray}
R = \left( \frac{\nu_{\rm m A}}{\nu_{\rm m B}} \right)^{2.1}.
\end{eqnarray}
Even if the spectral peak is caused by SSA, beyond our assumption, $R$ represents asymmetry in terms of SSA.
The spectrum of power-law synchrotron radiation with SSA is expressed by
\begin{eqnarray}
S_{\nu} = S_0 \nu^{2.5} \left[1 -  \exp(-\tau_{\rm s} \nu^{\alpha_0 - 2.5}) \right], \label{eqn:ssaspectrum}
\end{eqnarray}
where $\tau_s$ is the SSA opacity coefficient.
The spectral peak $\nu_{\rm m}$ approximates to $\sim \tau_{\rm s}^{\frac{1}{2.5 - \alpha_0}}$.
Then the relation between $R$ and $\tau_{\rm s}$ will be
\begin{eqnarray}
R = \left( \frac{\tau_{\rm s A}}{\tau_{\rm s B}} \right)^{\frac{2.1}{2.5 - \alpha_0}}, \label{eqn:ssaasymmetry}
\end{eqnarray}
though $R$ is derived from the spectral fits using the FFA model.

\begin{figure}[htb]
\begin{center}
\psfig{file=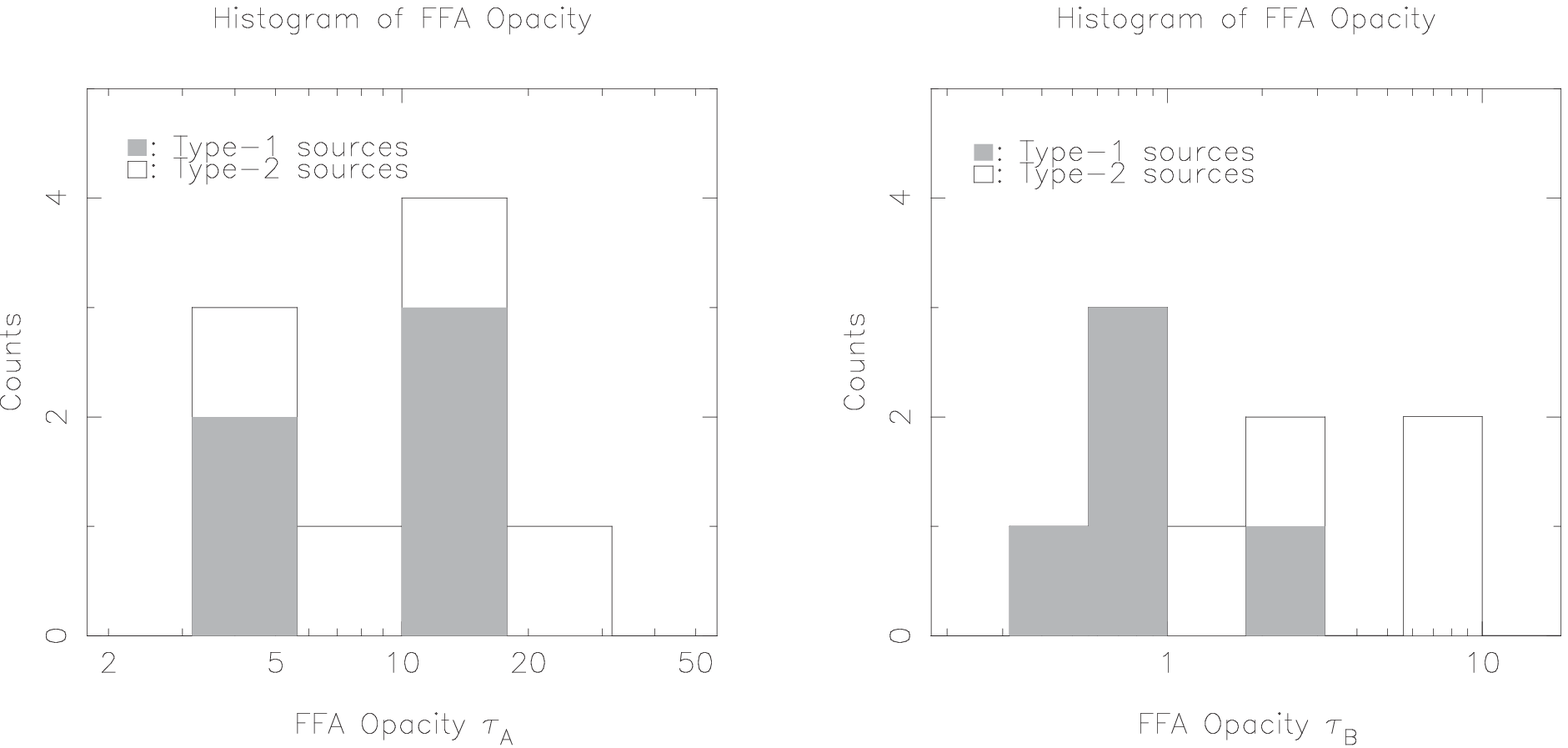,width=120mm}
\end{center}
\caption{\footnotesize Histograms of the FFA opacities towards components A ($\tau_{\rm fA}$: {\it left}) and B ($\tau_{\rm fB}$: {\it right}). 
Open and filled areas indicate type 2 (radio and Seyfert galaxies) and 1 (quasars), respectively.
Component B shows significant difference between type-1 and -2 sources, which component A does not.
}
\label{fig:FFAhistogram}
\end{figure}

The histogram of $R$ clearly exhibits the difference between type-1 and -2 sources (see figure \ref{fig:FFAhistogram}).  
The weighted means of $R$ for type-1 and -2 are $4.97 \pm 1.15$ and $1.36 \pm 0.51$, respectively.
Here, we perform a statistical test, whether any significant difference arises between the FFA opacity ratios of type-1 and -2 groups.
Let us put a testing hypothesis, which assumes $R$ is identical for these two subsets. 
Then, the $T$ value defined as
\begin{eqnarray}
T = \frac{|\bar{X_1} - \bar{X_2}|}{\sqrt{ \left( \frac{1}{n_1} + \frac{1}{n_2} \right) \left( \frac{S_1 + S_2}{n_1 + n_2 - 2} \right) }},
\end{eqnarray}
must follow the $t$ distribution with the degrees of freedom of 7.
Here, $\bar{X_1}$ and $\bar{X_2}$ are mean values of a variable $X$ in subsets 1 and 2, respectively, $n_1$ and $n_2$ are the number of data; $S_1$ and $S_2$ are the sum of residual squared.
The calculated value $T= 3.90 > t(7, 0.01) = 3.5$ rules out the testing hypothesis.
Thus, the opacity ratio of type-1 sources is significantly larger than that of type-2 sources, with the confidence larger than 99 \%.
Even if we take the T-test for logarithms $\ln R$, the values are $1.70 \pm 0.36$ and $0.57 \pm 0.52$ for type-1 and -2 sources, respectively.
Then, $T = 4.26 > t(7, 0.01)$ suggests, too, that type-1 sources are more asymmetric than type-2 sources in terms of FFA opacity.

The result is consistent with Barthel's unified scheme between RGs and QSOs (Barthel 1989).
If the line of sight is close to the jet axis, as thought to be type-1, a large difference of the path length in external plasma towards the lobes causes an asymmetric FFA. 
In case of type-2 sources, the line of sight is nearly perpendicular to the jet axis, so that a small difference in the path length results in a relatively symmetric FFA. 

To check if any bias is included in the statistics, we also test the distribution of intrinsic spectral indices $\alpha_0$ and ratios of intrinsic flux densities $FR = \frac{S_{0 \rm A}}{S_{0 \rm B}}$.
The spectral indices are $-0.64 \pm 0.16$ and $-1.12 \pm 0.25$ for type 1 and 2, respectively.
$T=3.06$ between $t(7,0.05) = 2.365$ and $t(7, 0.01) = 3.5$ suggests that $\alpha_0$ is likely to be different between the two classes.
The flux density ratios $\ln FR$ are $1.76 \pm 0.85$ and $0.17 \pm 0.62$ for type 1 and 2, respectively, and $T=2.74 < t(7, 0.1)$ indicate no significant difference.

The larger $\alpha_0$ for type-1 objects probably indicates contamination of the core component.
In fact, the flat spectrum of component A in 0646+600 and 1333+459 implies that these components are the core.
Alternatively, let us use $\alpha_0$ of component B (lobes or jets) for these two objects, i.e., $\alpha_0 = -1.54$ and $-1.45$ for 0646+600 and 1333+459, respectively, as are derived in {\it Case 2}.
The mean $\alpha_0$ for type-1 objects will be $-0.95 \pm 0.47$, which attributes to no significant difference between that of type-2 objects with $T = 0.59 < t(7, 0.1)$.

From the above statistical considerations we conclude that the two groups show no intrinsic difference, but are apparently different in terms of FFA opacities. 
What does this mean?
Here, we discuss the reason of the apparent difference of the opacity ratios.

\bigskip
{\it Case 1: The viewing angle of the jet in type-1 source is smaller than that of type-2 sources}

\begin{figure}[htb]
\begin{center}
\psfig{file=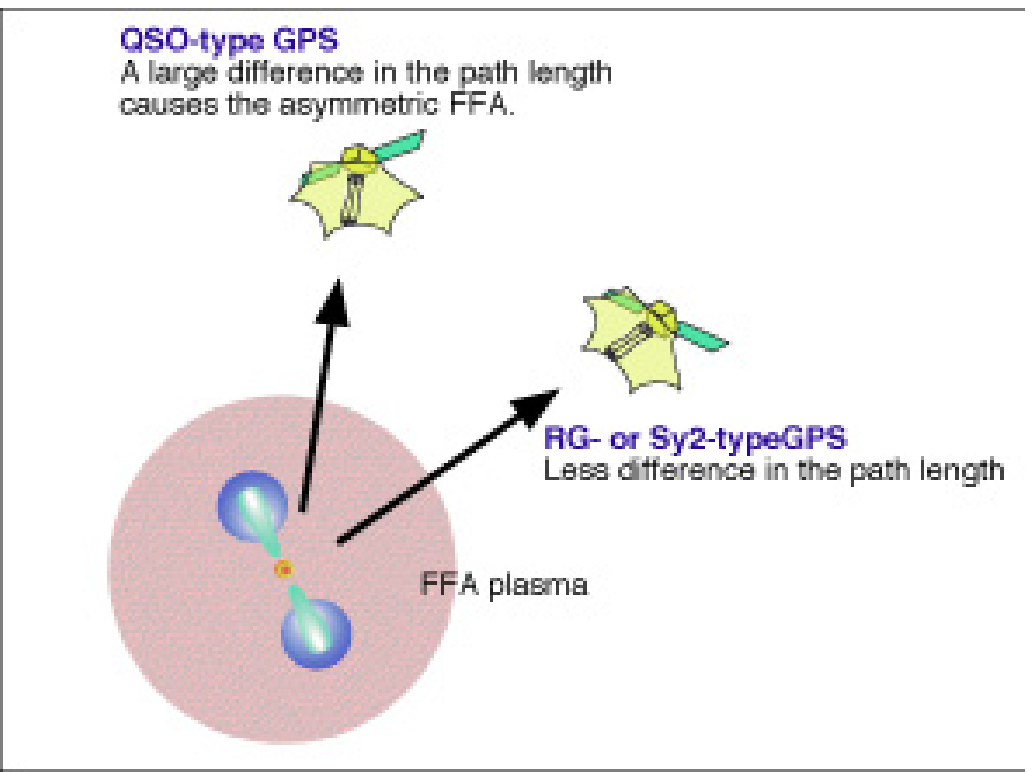,width=120mm}
\end{center}
\caption{\footnotesize A schematic diagram of a GPS source.
If the line of sight is close to the jet axis (type 1 sources), a large difference in the path length through the ambient plasma causes asymmetric FFA opacities.
A type 2 source, on the contrary, has small difference in the path length, since the line of sight is nearly perpendicular to the jet axis.
}
\label{fig:unifiedmodel}
\end{figure}

\bigskip
This idea is consistent with the unified scheme (Barthel 1989).
It simply explains why the FFA opacities are asymmetric towards type 1s, while no difference in intrinsic properties between type 1 and 2 is found.
The significant difference of the $\tau_{\rm fB}$ between type-1 and -2 objects (see figure \ref{fig:FFAhistogram}) can be understood in this model: the approaching component would have less path length in the ambient FFA plasma as shown in figure \ref{fig:unifiedmodel}.
The external absorption model is self-consistent with the FFA fitting.
This model does not require that SSA is intrinsically asymmetric towards the lobes.

\bigskip
{\it Case 2: Type-1 sources consist of the core and one-sided jet}

\begin{figure}[htb]
\begin{center}
\psfig{file=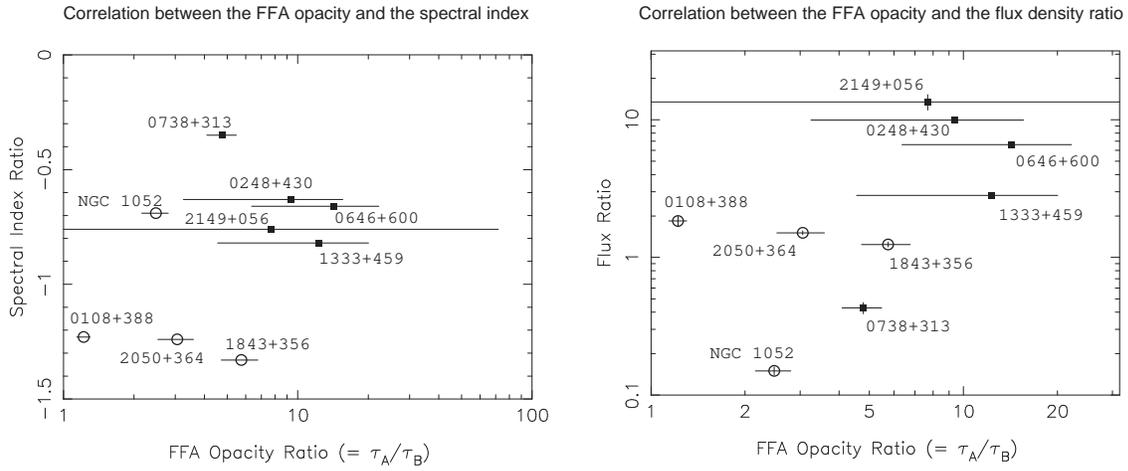,width=150mm}
\end{center}
\caption{\footnotesize {\it (left)}: Comparison between the FFA opacity ratio $R$ and intrinsic spectral index $\alpha_0$.
Open circles and filled squares stand for the type 2 and 1 objects, respectively.
Type-1 sources are likely to have a larger spectral index than that of type-2 objects.
A small correlation is found between $R$ and $\alpha_0$ with the normalized correlation coefficient of $\rho = 0.54$, which yields confidence limits of 87\%.
{\it (right)}: Comparison between the FFA asymmetry index $R$ and the flux density asymmetry index defined as $FR = \frac{S_{0 \rm A}}{S_{0 \rm B}}$.
There is No significant difference between the flux density asymmetry indices of type-1 and -2 objects.
No significant correlation is found between $R$ and $FR$ with the normalized correlation coefficient of $\rho = 0.29$.}
\label{fig:FFAcorrelation}
\end{figure}

\bigskip
In this case, the inverted spectrum towards the core component is likely due to SSA.
The intrinsic spectral index $\alpha_0$ of a core could be larger than that of a jet or a lobe.
As shown in figure \ref{fig:FFAcorrelation}, spectral indices of type-1 sources are likely to be larger than those of type-2 sources.
For example, the spectral fits for 0646+600 and 1333+459 will be much better if a larger $\alpha_0$ is put for components A.
In the case of 0646+600, $\alpha_0 = -0.11$ and $-1.54$ for components A and B result in the best fit, though we lose degrees of freedom, and we have $S_{0 \rm A} = 1.19 \pm 0.02$, $\tau_{\rm fA} = 6.22 \pm 0.25$, $S_{0 \rm B} = 8.32 \pm 0.14$ and $\tau_{\rm fB} = 9.91 \pm 0.25$.
In the case of 1333+459, $\alpha_0 = -0.50$ and $-1.45$ for A and B, respectively, are the best fit to derive $S_{0 \rm A} = 1.01 \pm 0.02$, $\tau_{\rm fA} = 7.7 \pm 0.3$, $S_{0 \rm B} = 4.89 \pm 0.11$ and $\tau_{\rm fB} = 7.8 \pm 0.3$.
In both cases, components A are likely to be a core, rather than a lobe, because of their flat spectrum.
If we assume intrinsic bipolarity of the jet, it is necessary to consider why the counterjet is unseen.

One may consider that the Doppler boosting effect possibly causes the apparent one-sided jet.
In this interpretation, component B is approaching towards us to be amplified, while the unseen counterjet is receding to be dimmed.
However, the Doppler boosting effect often results in variability of flux densities and a large polarization degree.
These properties are unlikely for GPS sources.

An alternative interpretation is that the unseen counterjet is severely obscured via FFA.
This could be an attractive model in which the diffuse emissions opposite to components B in 0646+600 and 0738+313 can be understood as the counterjets.
Deeper imaging capabilities at higher frequencies are necessary to confirm the counterjet.

Whatever attenuates the counterjet, a smaller viewing angle in type-1 sources than that of type-2 sources is implied.
Thus, this case is also consistent with the Barthel's unified scheme.
The path length in the external absorber towards component B of type-1 sources is expected to be shorter than those of type-2 sources (see figure \ref{fig:unifiedmodel}). 
Smaller $\tau_{\rm fB}$ of type-1 sources than that of type-2 sources can be understood in this context.

\bigskip
{\it Case 3: Type-1 sources are smaller than type-2 sources.}

\bigskip
When the plasma density decreases as a function of the radius from the nucleus, like the isothermal King model (King 1972), denser FFA opacities near the center will be produced.
The asymmetry of the opacity can be enhanced, even if the viewing angle remains the same, if the source size is small (Kameno et al. 2001).
This idea is similar to the hypothesis that the FFA plasma is denser towards type-1 sources than towards type-2 sources, as is scaled by the core radius of the ambient FFA plasma.
However, this model requires that the mean opacity of type-1 sources should be larger than that of type-2 sources.
The histogram $\tau_{\rm B}$ (figure \ref{fig:FFAhistogram}) shows the opposite behavior, so that this idea is not supported.

\begin{figure}[htb]
\begin{center}
\psfig{file=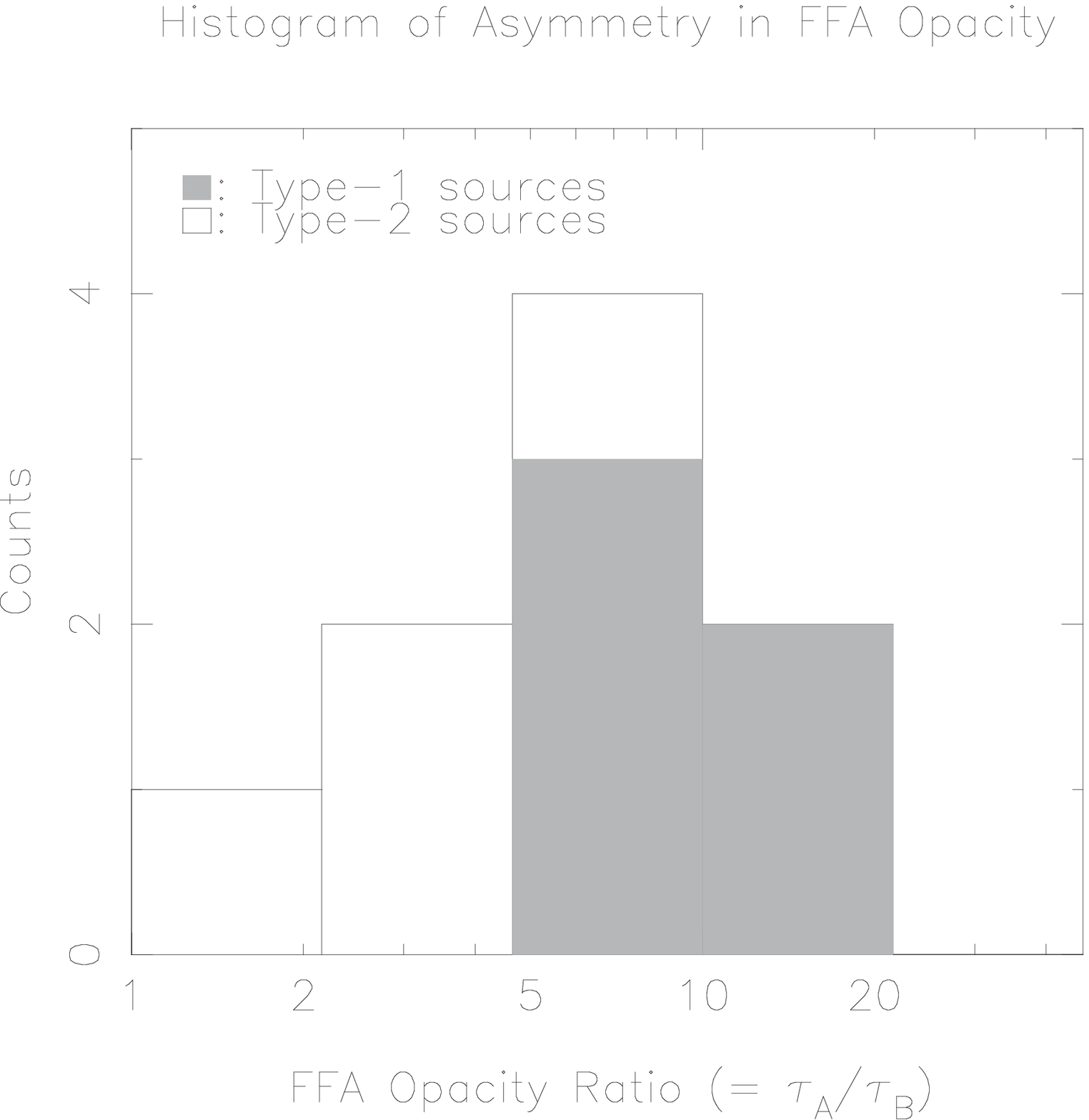,width=120mm}
\end{center}
\caption{\footnotesize Histogram of the FFA opacity ratio $R$ defined as $R = \frac{\tau_{\rm fA}}{\tau_{\rm fB}}$, where $\tau_{\rm fA}$ and $\tau_{\rm fB}$ are the FFA coefficients of double component.
Since components are labeled in order of $\tau_{\rm f}$, $R$ is always larger than or equal to 1.
Open and filled areas indicate type 2 (radio and Seyfert galaxies) and 1 (quasars), respectively.
This histogram shows that type 1 sources are significantly asymmetric than type 2 sources are, in terms of FFA opacities.
}
\label{fig:FFAasymmetry}
\end{figure}

\bigskip
Consequently, the simplest explanation for the difference in the asymmetry of opacity is the difference of viewing angle.
This should be discussed coupled with the absorption mechanisms, FFA or SSA.
If SSA dominated in the low-frequency cutoff, we expect that the peak frequencies of components A and B would be statistically unbiased in the co-moving frame with each synchrotron emitter.
When the synchrotron sources emanate from the core at a relativistic speed, the peak frequencies of approaching and receding components would become higher and lower, respectively, due to the Doppler effect.
Together, the flux densities of approaching and receding components would be also amplified and dimmed, respectively, by the Doppler effect.
Hence, the flux density ratio $FR = S_{0\rm A} / S_{0\rm B}$ is expected to correlate with the peak frequency.
However, no significant correlation is found between $R = \tau_{\rm fA} / \tau_{\rm fB}$ which is related to the peak frequency, and $FR$, with the correlation coefficient of $\rho = 0.29$ (see figure \ref{fig:FFAcorrelation}).
These statistics of $R$ and $FR$ does not support the simple SSA-only model, and requires that double lobes are intrinsically asymmetric.
Saikia et al. (1995) reported that compact steep spectrum sources with small linear size are intrinsically asymmetric.
Carvalho (1998) showed that interaction of the jet with a non-homogeneous intragalactic medium can result in asymmetric evolution.
Since the peak frequency in terms of SSA is anti-correlated with the linear size (O'Dea et al. 1998), the intrinsic asymmetry of the SSA opacity can be produced.
The presence of SSA and its intrinsic asymmetry cannot be ruled out, nevertheless, the unified scheme coupled with the FFA model can simply account for the difference in asymmetry of opacities.

\section{Conclusions}

VLBA observations for nine sources at 2.3, 8.4, and 15.4 GHz have been carried out, to reveal the morphologies of all objects.
Spectral model fitting is applied to obtain spatial distribution of FFA opacities towards the radio emission of individual sources.
A difference between type-1 (quasars and Seyfert 1 galaxies) and type-2 (radio and Seyfert 2 galaxies) is found, in terms of the FFA opacity ratio $R = \tau_{\rm fA} / \tau_{\rm fB}$ between two components A and B of each object.
Type-1 objects tend to show significantly larger opacity ratios than type-2 sources do.
Asymmetry in FFA opacities suggests that path lengths through ambient absorbers towards twin lobes are significantly different.
Therefore, larger opacity ratios of type 1 objects indicate the axes between lobes are smaller than those of type 2 objects (see figure \ref{fig:unifiedmodel}).
This result supports the unified scheme between quasars and radio galaxies proposed by Barthel (1989).


%
%




\section*{Acknowledgments}

We the VLBA operated by the National Radio Astronomy Observatory, which is a facility of the National Science Foundation (NSF).
This work was supported by Grant in Aid (c) 13640248 from Japan Society for the Promotion of Science.

\section*{References}






\reference Antonucci, R. R. J., \& Miller, J. S. \  1985, ApJ, 197, 621

\reference Barthel, P. D. \ 1989, ApJ, 336, 606

\reference Bicknell, G. V., Dopita, M. A., \& O'Dea, C. P. \  1997, ApJ, 485, 112

\reference Carvalho, J. C. \ 1998, A\&A, 329, 845

\reference Kameno, S, Horiuchi, S., Shen, Z.-Q., Inoue, M., Kobayashi, H., Hirabayashi, H., \& Murata, Y. \  2000, PASJ, 52, 209

\reference King, I. R. \ 1972, ApJ. 174, L123

\reference Kameno, S., Sawada-Satoh, S., Inoue, M., Shen, Z.-Q., \& Wajima, K. \  2001, PASJ, 53, 169

\reference de Vries, W. H., Barthel, P. D., \& O'Dea, C. P.\  1997, A\&A, 321, 105

\reference Miller, J. S., \& Antonucci, R. R. J.\  1983, ApJ, 271, L7

\reference O'Dea C. P. \  1998, PASP, 110, 493


\end{document}